\documentclass[preprint,12pt]{elsarticle}

\usepackage{amssymb}
\usepackage{amsmath}
\usepackage{amsfonts}
\usepackage{graphicx}
\usepackage{natbib}
\usepackage{textcomp}
\usepackage{color}
\usepackage{array,tabularx}
\def\<{\langle}
\def\>{\rangle}
\def\gpc{g_{\rm pc}}

\journal{Physica A}

\begin{document}

\begin{frontmatter}

%% Title, authors and addresses

%% use the tnoteref command within \title for footnotes;
%% use the tnotetext command for theassociated footnote;
%% use the fnref command within \author or \address for footnotes;
%% use the fntext command for theassociated footnote;
%% use the corref command within \author for corresponding author footnotes;
%% use the cortext command for theassociated footnote;
%% use the ead command for the email address,
%% and the form \ead[url] for the home page:
%% \title{Title\tnoteref{label1}}
%% \tnotetext[label1]{}
%% \author{Name\corref{cor1}\fnref{label2}}
%% \ead{email address}
%% \ead[url]{home page}
%% \fntext[label2]{}
%% \cortext[cor1]{}
%% \address{Address\fnref{label3}}
%% \fntext[label3]{}

%\title{\textcolor{red}{Mixed first order percolation and jamming transition}}
\title{Jamming as a random first-order percolation transition}

%% use optional labels to link authors explicitly to addresses:
\author[label1,label3]{Antonio Piscitelli}
\author[label1,label2]{Antonio Coniglio}
\author[label1]{Annalisa Fierro}
\author[label1,label3]{Massimo Pica Ciamarra}
\address[label1]{CNR-SPIN, c/o Complesso di Monte S. Angelo, via Cinthia - 80126 - Napoli, Italy}
\address[label2]{Physics Department, Universit\`a degli Studi di Napoli “Federico II”, Napoli, Italy}
\address[label3]{Division of Physics and Applied Physics, School of Physical and Mathematical Sciences, Nanyang Technological University, Singapore}

%\author{Antonio Coniglio, Annalisa Fierro, Massimo Pica Ciamarra, Antonio Piscitelli}

%\address{}

\begin{abstract}
We determine the dimensional dependence of the percolative exponents of the jamming transition via numerical simulations in four and five spatial dimensions. 
These novel results complement literature ones, and establish jamming as a mixed first-order percolation transition, with critical exponents $\beta =0$, $\gamma = 2$, $\alpha = 0$ and the finite size scaling exponent $\nu^* = 2/d$ for values of the spatial dimension $d \geq 2$. We argue that the upper critical dimension is $d_u=2$ and the connectedness length exponent is $\nu =1$.
\end{abstract}

%\begin{keyword}
%% keywords here, in the form: keyword \sep keyword

%% PACS codes here, in the form: \PACS code \sep code

%% MSC codes here, in the form: \MSC code \sep code
%% or \MSC[2008] code \sep code (2000 is the default)

%\end{keyword}

\end{frontmatter}

%% \linenumbers

%% main text
\section{Introduction}
\label{intro}
Granular materials, emulsions, foams, and all systems of particles interacting via finite ranged repulsive interactions, for which thermal motion is negligible, exhibit a jamming transition on increasing the volume fraction, signalling the onset of mechanical rigidity~\cite{nu4}.
The transition has a first-order character, when approached from the unjammed phase, as the density of jammed particles exhibits a jump, from zero to a macroscopic value.
Equivalently, the average contact number  per particle, $Z$, jumps discontinuously from $Z = 0$, in the unjammed phase, to the isostatic value $Z_{\rm iso}=2d$.

The approach to jamming naturally lends itself to a percolative description, with control parameter $\epsilon = |(\phi-\phi_j)/\phi_j|$, where $\phi$ is the fraction of the total volume occupied by the particles and $\phi_j$ is the jamming threshold.
The mixed character of the transition, however, made difficult the identification of the percolative critical exponents, and the design of simple models~\cite{schwarz_onset_2006,jeng_force-balance_2010} able to capture the jamming phenomenology.
In percolation theory~\cite{stauffer,havlin,Stauffer82}, the density of the percolating cluster, which is the order parameter, scales as $P \propto \epsilon^\beta$ for $\epsilon > 0$ (above the transition), while, below  the transition, $P = 0$. 
The mean cluster size $S = \sum s^2 n(s)/\sum sn(s)$, where $n(s)$ is the number of clusters of size $s$, scales as $S \propto \left| \epsilon \right|^{-\gamma}$. 
%The same exponent controls the divergence of the fluctuations of the size of the percolating cluster~\cite{Coniglio1980}, $\Delta P^2 \propto \left| \epsilon \right|^{-\gamma}$.
The pair connected correlation function in $d$ spatial dimensions is $\gpc(r) \propto r^{-d+2-\eta} f(-r/\xi)$, with $f(x)$ being an exponential decreasing function for large $x$  and the connectedness length diverging at the percolation transition as $\xi \propto \left| \epsilon \right|^{-\nu}$. 
The fractal dimension of the cluster is $D=d-\beta/\nu$.
Different set of exponents, which satisfy 
the scaling and hyperscaling law $2\beta + \gamma = 2-\alpha = d\nu$, identify different universality classes. 
More precisely hyperscaling relations contain the space dimension $d$.
While scaling relations are verified for all dimensions $d$, hyperscaling do not hold for dimensions greater than the upper critical dimension $d_u$, where mean field exponents hold. 

To investigate the percolative character of the jamming transition, we consider a soft sphere packing at zero temperature prepared according to some protocol. 
Since the number of touching spheres is zero below the threshold and becomes finite at the threshold, the percolation order parameter $P$, defined as the density of touching spheres in the spanning cluster, jumps from zero to a finite value $P_c$ at the transition~\cite{nu2,nu3}. 
This discontinuity suggests $\beta = 0$. 
%The study of other percolative quantities is more strictly related to the chosen notion of connectivity.
%For instance, if particles are considered connected when in contact, then below the transition all clusters have size one, as there are no interparticle contacts. 
%Hence, one would find $\xi$ constant and $\nu = 0$.
Finite size scaling investigations allow to extract an exponent $\nu^*$ which below or at the upper critical dimensionality $d_u$ coincides with the connectedness length exponent $\nu$. Conversely, for $d>d_u$, while $\nu$ is given by its mean field value independent on the dimensionality, $\nu^*$ depends on the dimensionality~\cite{binder,WY} in such a way that using $\nu^*$ instead of $\nu$, the hyperscaling relation are satisfied also above $d_u$. 
The precise value of $\nu^*$ is however debated~\cite{nu4,nu2,nu3,nu1,aste}.
In particular, earlier numerical works gave $\nu^* \simeq 0.71\pm 0.08$, in both $d = 2$ and $d = 3$ spatial dimensions, suggesting a value $\nu^* = 2/3$ independent on the dimensionality \cite{nu2,nu3}, while a more recent investigation \cite{aste} gives
$\nu^* \simeq 1 $ for $d=2$ and $\nu^*  \simeq 0.66$ for $d=3$, suggesting that the exponent $\nu^*$ might depend on the spatial dimensionality as $\nu^* = 2/d$.

To set this issue, here we consider the case for $d =4$ and $d=5$. 
We have performed large scale simulations and an accurate analysis, which strongly support $\nu^*=2/d$ also for $d=4$ and $d=5$. 
We present these results in Sect.~\ref{sec:results}.
Interestingly the value $\nu^* = 2/d$ is that suggested for the ideal glass transition by Kirkpatrick, Tirumalai and Wolyness within the RFOT theory~\cite{KTW}. 
In Sect.~\ref{sec:rfopt} we describe jamming as random first-order percolation transition.
In Sect.~\ref{sec:1d}  we show explicitly that jamming of hard spheres in $1d$ is described by such percolative model although with different exponents.

\section{Numerical results~\label{sec:results}}
\subsection{Numerical protocol}
We have developed a molecular dynamics code to simulate systems of Harmonic spheres in arbitrary dimensions.
We consider monodisperse systems of sphere of radius $R$ and volume $v_d = \frac{\pi^{d/2}R^d}{\Gamma(d/2+1)}$, only interacting if their separation $r$ is smaller than $2R$, with energy given by $V(r) = \epsilon (r-2R)^2$.
To investigate the jamming transition, we randomly place $N$ spheres in a hypercube of volume $L^d$, where $L$ fixes the volume fraction $\phi=N v_d/L^d$.
We then minimize the energy of the system combining the conjugate-gradient minimization algorithm and a damped dynamics, and record properties of the final configuration.
The final value of the energy per particle after the minimization, $E$, allows us to distinguish between jammed configuration, $E > E_t$, and unjammed ones, $E < E_t$, with $E_t$ a threshold value. 
Here, we fix $E_t = 10^{-12}\epsilon$ but have checked that the results are insensitive to this choice.

As in $d=2$ and $d=3$, also for $d = 4$ and $d = 5$ the order parameter $P$, the fraction of particles of the percolating cluster made of touching spheres, jumps from zero to a finite value.  Hence, the percolative exponent associated with the behaviour of the order parameter is $\beta = 0$, for all dimensions investigated and is expected to be zero for any dimensions.
At the same time, the mean contact number is zero below the transition and jumps to the isostatic value at the transition $Z_{\rm iso}$.
Specifically, we have found $Z-Z_{\rm iso} \propto p$, with $p$ the pressure. 

\begin{figure}[!!t]
\begin{center}
\includegraphics*[scale=0.55]{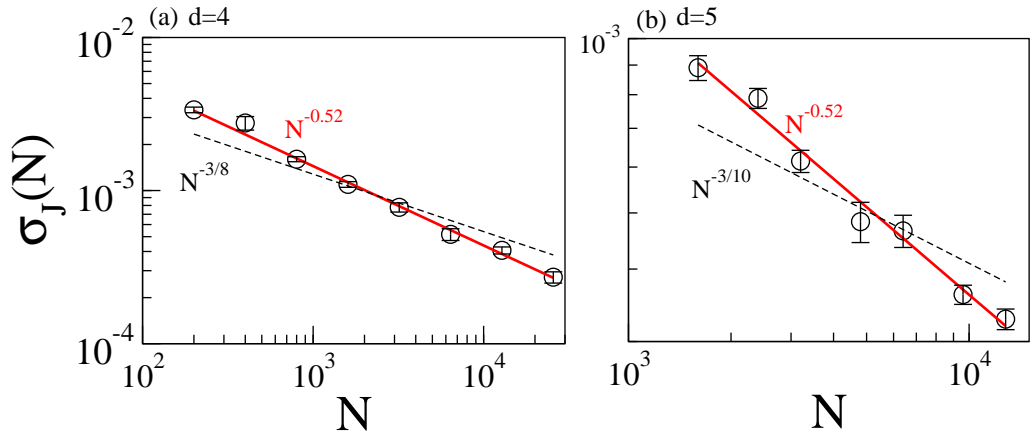}  
\end{center}
\caption{\label{fig:sigma} 
Size dependence of the jamming probability standard deviation $\sigma_J$ (see ~Eq.(\ref{eq:pj})) in $d = 4$ (a) and $d = 5$ (b). Data are compatible with $N^{-\Omega}$ with $\Omega = 1/2$, implying $\nu^* = 1/\Omega d= 2/d$.
}
\end{figure}
Repeating the minimization procedure $100$ times, starting from different random configurations, we associate to each system size and volume fraction a jamming probability $P_J(\phi,N)$, which is the fraction of our minimizations ending in a jammed state.
The jamming probability vanishes for small volume fraction, while it approaches $1$ for large volume fraction.
As observed in both $2d$ and $3d$~\cite{aste,pica_ciamarra_disordered_2010}, also in $4d$ and $5d$ the volume fraction dependence is well described by an error function,
\begin{equation}
P_J(\phi,N) = \frac{1}{2}\left[1+{\rm erf}\left(\frac{\phi-\phi_J(N)}{\sigma_J(N) \sqrt{2}}\right) \right]. 
\label{eq:pj}
\end{equation}
Hence, each system size $N$ is characterized by a typical jamming volume fraction, $\phi_J(N)$, and standard deviation, $\sigma_J(N)$, we extract from a numerical fit of the $P_J(\phi,N)$ data.

We expect~\cite{nu2,nu3}  $\sigma_J(N)=\sigma_0 L^{-1/\nu^*}$, which gives $\sigma_J(N) \propto N^{-\Omega}$ with $\Omega = 1/d \nu^*$, since  the system size is $L \propto N^{1/d}$. In Ref.~\cite{aste}, the value of $\Omega=0.5$ was obtained as fitting parameter in both $d=2$ and $3$, giving $\nu^*=1$ and $2/3$, respectively. These findings suggest $\nu^*=2/d$ for all values of $d$. However, such predictions are non easily discernible from the value $\nu^*=2/3$ obtained by previous investigations~\cite{nu2,nu3} in both $d = 2$ and $3$.

In Fig.~\ref{fig:sigma}a and Fig.~\ref{fig:sigma}b the standard deviation $\sigma_J(N)$ is plotted as function of $N$ for $d=4$ and $d=5$, respectively. The best fit of $\sigma_J(N)$ as $N^{-\Omega}$ gives  $\Omega = 0.52 \pm 0.01$ in both cases, whereas the values $\Omega \simeq 3/8$ and $3/10$, obtained from $\nu^*=2/3$, do not well described our data, as we directly demonstrate in figures.

\begin{figure}[!!t]
\begin{center}
\includegraphics*[scale=0.55]{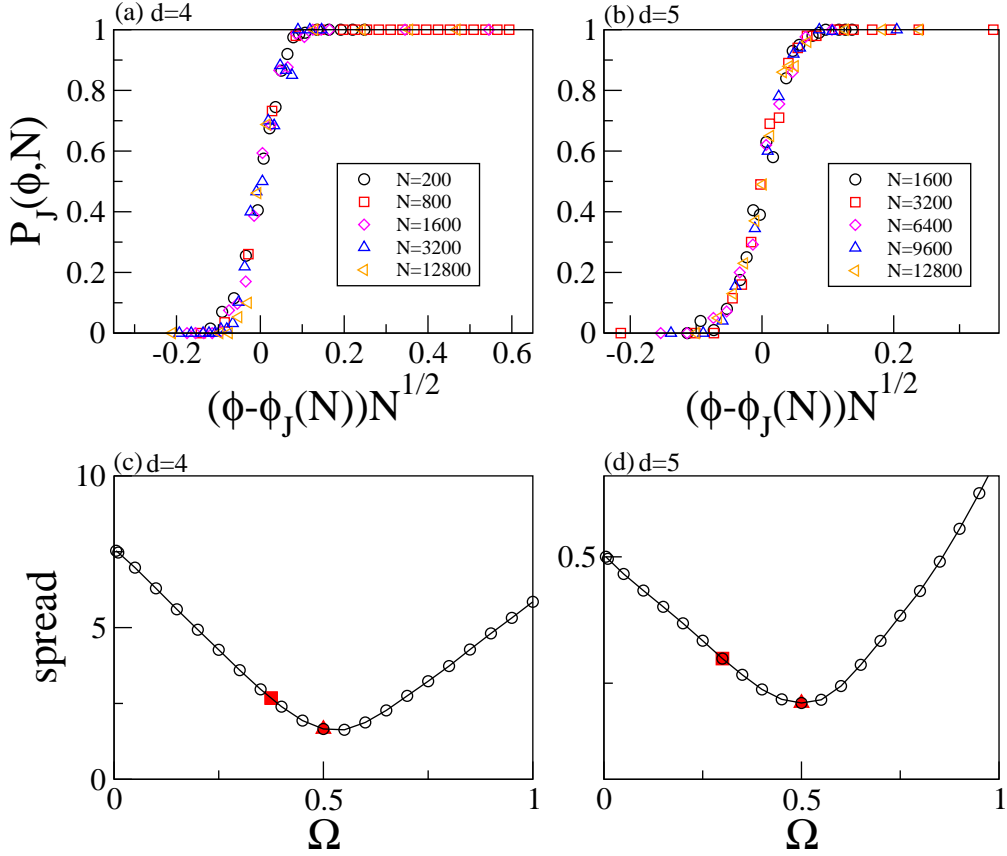}     
\end{center}
\caption{\label{fig:scaling} The jamming probabilities fall on the same master curve when plotted versus $(\phi-\phi_J(N))N^\Omega$ with $\Omega = 1/2$, in both $d=4$ (a) and $d=5$ (b).
The dependence of the spreading observed in the collapse on the exponent $\Omega$, confirms that $\Omega = 1/2$ provides the better collapse in both $d=4$ (c) and $d=5$ (d).}
\end{figure}

To further support the value $\nu^* = 2/d$, we show in Fig.~\ref{fig:scaling}(a) ($d=4$) and (b) ($d = 5$) that data for the jamming probability for different $N$ collapse on a master curve, when plotted as a function of $(\phi-\phi_J(N))N^\Omega$, with $\Omega = 1/2$. 

To investigate whether this $\Omega$ value is the one leading to the best data collapse, we evaluate a spread function  defined as
\begin{equation}
    {\rm spread}(\Omega) = \frac{1}{n_c} \sum_{N_1>N_2} \int \left|P_J(x,N_1)-P_J(x,N_2)\right| dx,
\end{equation}
where $n_c$ is the number of distinct $N_1,N_2$ couples, and $x(\Omega) = (\phi-\phi_J(N))N^{\Omega}$.
Smaller values of the spread indicate a better data collapse.
Fig.~\ref{fig:scaling}(c) and (d) illustrate the dependence of the spread on $\Omega$, in $d = 4$ and $d = 5$.
Regardless of the dimensionality, we found the spread to have a minimum at $\Omega \simeq 0.5$ (red triangles).
The values $\Omega = 3/8$ and $3/10$ (red squares), suggested by $\nu^* = 2/3$, provide a worse data collapse.

Considering previous results~\cite{aste,pica_ciamarra_disordered_2010} in $d = 2$ and $d = 3$, and our present ones in $d = 4$ and $d = 5$, it appears that $\Omega \simeq 0.5$, regardless of the dimensionality.
This suggests  that the exponent $\nu^*$ depends on the dimensionality as $\nu^* = 2/d$.

%Having estimated $\nu^* = 2/d$, we can also estimate the exponent $\gamma$ as that controlling the divergence of the fluctuations of the size of the percolating cluster. 
%Since $\beta = 0$ the percolating cluster, when present, 
%has size $\propto L^d$. 
%Hence, the sample-to-sample fluctuations of the size of the percolating cluster $\Delta P^2 \propto |\epsilon|^{-\gamma}$ ~\cite{Coniglio1980} are
%$\Delta P^2 \propto L^d [P_J(\phi,N)-P_J^2(\phi,N)]
%\propto \xi^{d} f(\xi(\epsilon)/L)$, where $f(x)$ is a constant for $x \ll 1$, $f(x) \propto (\xi/L)^{-d}$ for $x \gg 1$. 
%Hence, we derive $\gamma = d\nu^* = 2$.

\section{Random first-order percolation transition~\label{sec:rfopt}}
We have found the approach to jamming to be a percolation transition characterised by a jump in the order parameter with an exponent $\beta=0$, a divergence in the finite size scaling length with an exponent $\nu^*=2/d$.%, and fluctuations of the size of the percolating cluster diverging with exponent $\gamma = 2$.
The critical exponents $\alpha$ and $\gamma$  can be obtained by scaling argument: $\alpha = 2-d \nu^* = 0$ and $\gamma=d\nu^*-2\beta=2-\alpha=2$. 
This transition has a mixed character, being intermediate between a first- and a second-order transition, as the random first-order transition introduced in the context of the glass transition.
Therefore, we name it random first-order percolation transition~\cite{KTW}.
Summarising, the exponents of this percolation transition are given by:
\begin{equation}
\label{exponents}
\beta =0,\qquad \gamma = 2,\qquad \alpha = 0,\qquad \nu^* = 2/d,
\end{equation}
and satisfy the scaling and hyperscaling law:
\begin{equation}
\label{scaling}
2\beta + \gamma = 2-\alpha = d\nu^*.
\end{equation}
Interestingly we have $\alpha=0$, $\beta=0$, $\gamma = 2$, which are independent on the dimensionality.  
The only exponent depending on the dimensionality is $\nu^*=2/d$. Although $\nu^*=2/d$ has been verified only for $d=2,3,4,5$ we expect that should be valid for all dimensions $d$.

We discuss now the question concerning the upper critical dimensionality $d_u$. It has been suggested in the literature ~\cite{nu2,ikeda}, that 
for jamming $d_u=2$.  Our result, that $\alpha$, $\beta$, $\gamma $ for $d \geq 2$ are independent on the dimensionality, strongly suggests that  they are mean field exponents and strongly supports $d_u=2$. At $d=d_u=2$, $\nu = \nu^* = 1$ and, since the value of the exponent at the upper critical dimensionality coincides with its mean field value, we have   $\nu =1$ for any $d \geq 2$.
In conclusion if $d_u=2$ is the upper critical dimension, the critical exponents for this random first-order percolation are, 
for any $d\geq 2$, $\alpha=0$, $\beta=0$, $\gamma = 2$, and $\nu=1$. These are mean field exponents, which obey scaling 
$\alpha+2\beta + \gamma = 2$ for any $d$ and  hyperscaling    $2\beta + \gamma = d\nu$ only for $d=2$. 
Hyperscaling is restored for any $d$ only if $\nu$ is replaced by $\nu^*$.

\begin{figure}[!!t]
\begin{center}
\includegraphics*[scale=0.5]{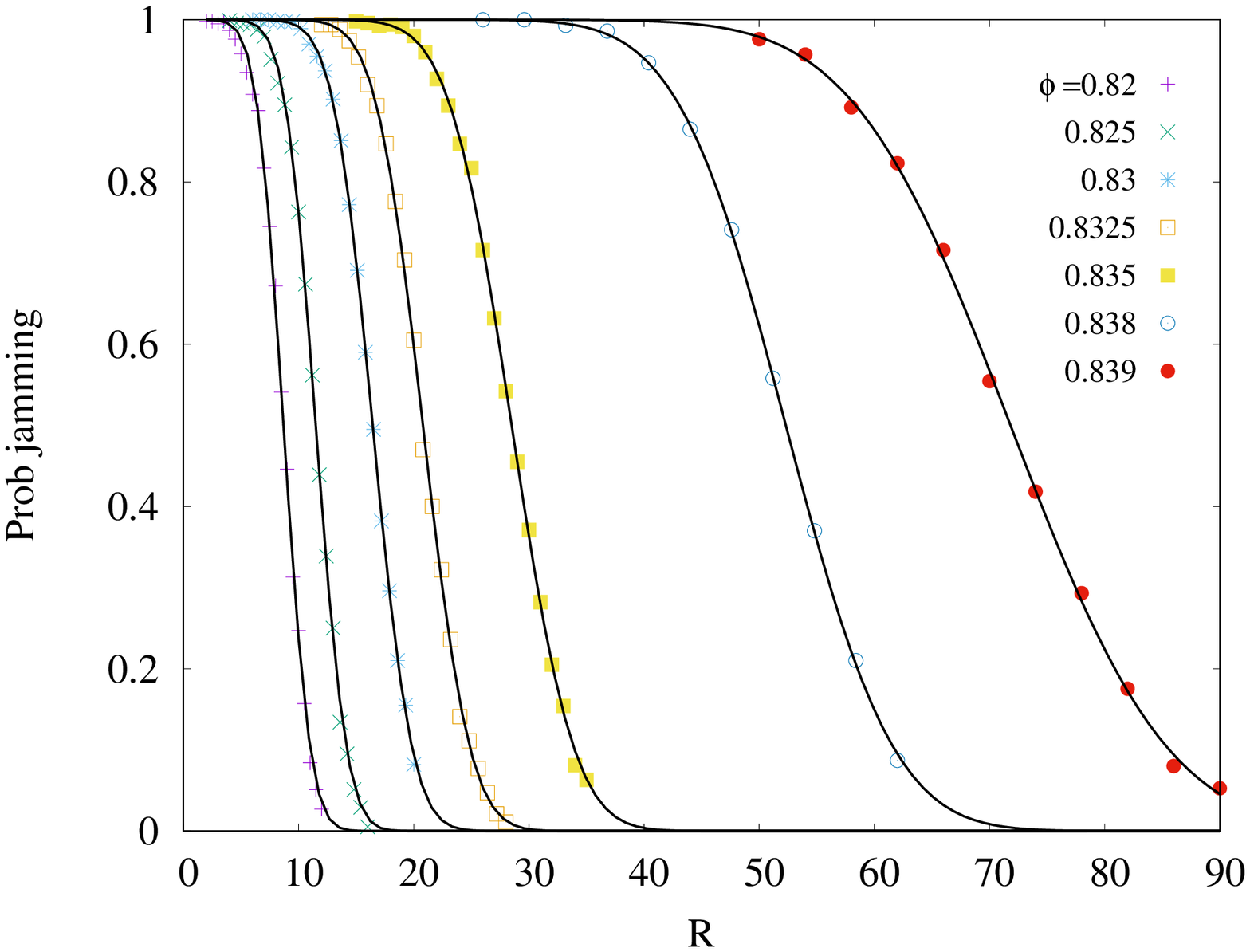}     
\includegraphics*[scale=0.5]{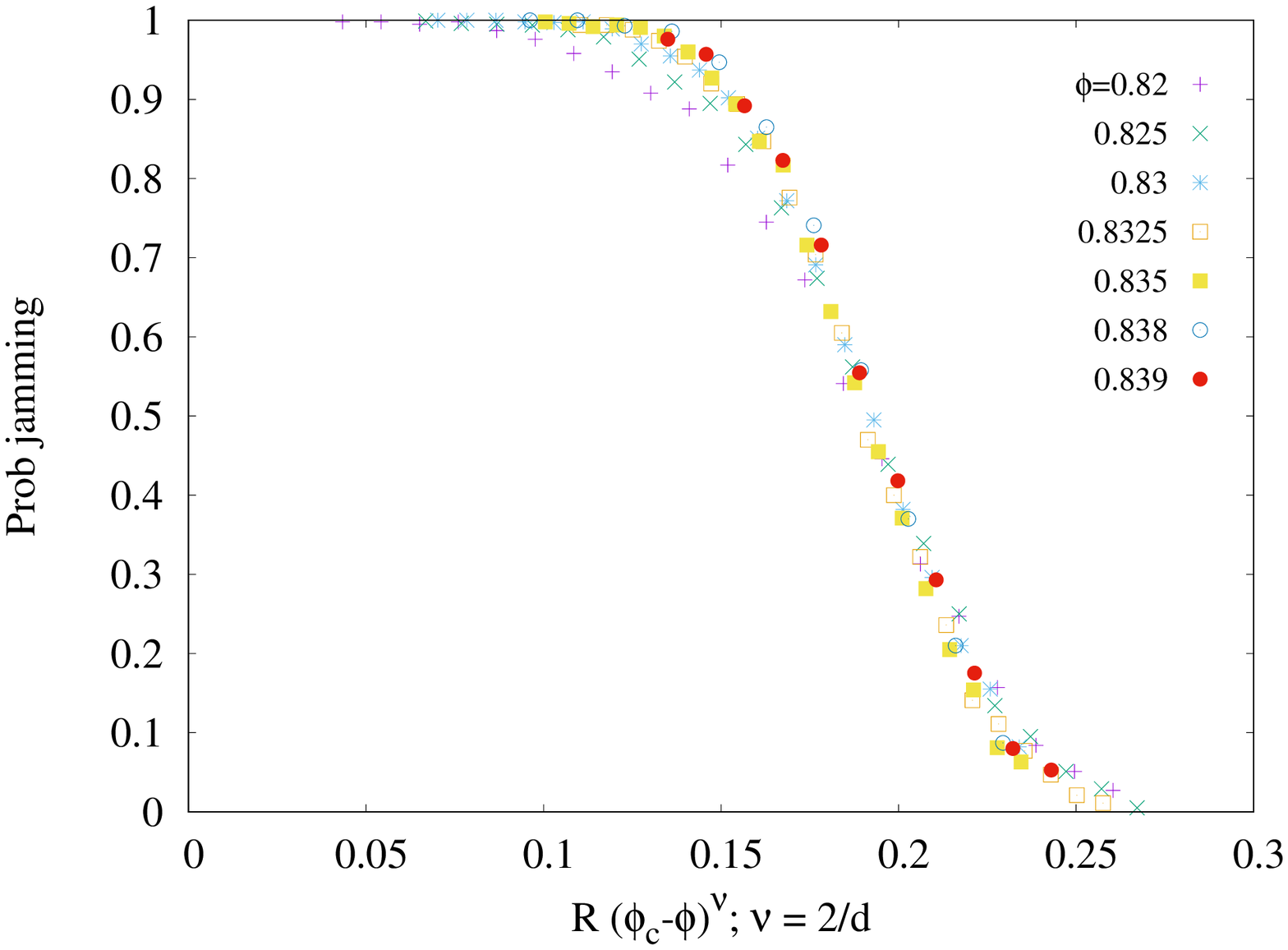}     
\end{center}
\caption{\label{fig:PTscaling} If the particles enclosed in a circle of radius $R$ of an unjammed configuration of volume fraction $\phi$ have their positions randomized within the circle, the subsequent energy minimization conducted keeping frozen the particles outside the circle may lead to a jammed or to an unjammed configuration. 
The top panel illustrates the probability that the resulting configuration is jammed as a function of $R$, for different values of the volume fraction. Each data point is the average of $10^3$ trials. 
The jamming probabilities function collapse on a master curve with $R$ is scaled by $R^*\propto (\phi_c-\phi)^{-\nu}$ and $\nu = 1$.
}
\end{figure}

%One may wonder which are the clusters below the jamming transition. Naively one may define a  cluster as a group of touching particles.  However, below jamming the coordination number $Z$ is zero, therefore accordingly there would be only one site clusters. 
 
We now show that the above percolative exponents are compatible with the existence of clusters of typical size $\xi$. 
Specifically, while in the next section we will give an appropriate definition of clusters for $1d$ hard spheres, 
 here we focus on $d = 2$.
%For dimension $d= 2$, due to topological disorder, the cluster definition below jamming is more elaborated. 
We propose a definition reminiscent of the point-to-set procedure used in the context of glass transition by Biroli et al.~\cite{biroli} to characterize the length associated to the cooperative rearranging regions introduced by Adam and Gibbs~\cite{AG}, and further elaborated in the theory of random first-order transition (RFOT) by Kirkpatrick Thirumalai and Woliness \cite{KTW}. 
For a fixed volume fraction $\phi$ below jamming, we consider a configuration of unjammed particles generated with the numerical protocol described in Sect. 2, an energy minimization starting from a random arrangement of particles.
We then focus on a set of particles enclosed in a circle of radius $R$, considering frozen the particles outside this circle.
We randomize the particles within the circle, ensuring that they remain within the circle, and then minimize the energy, always keeping the outer particles frozen.
As a result, the particles inside the circle might be both in a jammed and in an unjammed configuration.
We calculate the jamming probability of this set of particles, as a function of $R$ and the volume fraction $\phi$ . 
Such probability will be a decreasing function of $R$. 
Say $R^*$ the value of R corresponding to the jamming probability equal to $1/2$ . 
%This value of $R^*$ corresponds to the linear dimension of a typical cluster at volume fraction $\phi$. 
The set of jammed particles in the cavity of size $R^*$ defines a typical cluster at volume fraction $\phi$ with linear dimension $\xi=R^*$.
For large values of $R$ the procedure is not much different from the finite-size scaling approach. 
Therefore we expect that, as $\phi$ approaches the jamming threshold, $R^*$ diverges with the exponent $\nu = 1$. Numerical simulations in $2d$ show indeed that this is the case (see Fig.~\ref{fig:PTscaling}). Moreover, at the threshold, the critical cluster coincides with the jammed percolating one. For dimension $d>2$ the cluster definition is more elaborated, due to the presence of two scaling lengths. This issue will not be considered here.

We finally remark that this percolation transition is an example of explosive percolation, e.g. Ref.~\cite{Achlioptas, Araujo2010, Riordan} and references therein.

\section{Jamming in $1d$\label{sec:1d}}
In $d = 1$, jamming corresponds to a percolation problem which can be solved exactly. 
We first consider the random percolation model on a lattice~\cite{RSK}, where the percolation in $d=1$ has the properties of a random first-order percolation transition. 
Indeed the cluster size distribution is given by $n(s) = (1-p)^2p^s$, where $p$ is the probability that a site is occupied, the mean cluster size is given by $S=(1+p)/(1-p)$ diverging at the percolation threshold $p_c=1$ with an exponent $\gamma=1$. 
The density of sites in the spanning cluster is zero for $p<1$ and jumps to $1$ at the percolation threshold $p_c=1$ with the critical exponent $\beta =0$. The other critical exponents are given by $\alpha =1$, $\nu= 1$ and $\tau =2$.
%Note that the cluster size distribution $n(s)$ obeys the scaling law (\ref{ns}) being $s^* = \xi =| \ln p | \sim (1-p)$.

We consider now jamming of hard sphere model in $d = 1$. 
In this case, the density of percolating jammed sites is $0$ below the jamming transition and then jumps to $1$ at volume fraction $\phi_j=1$, where a macroscopic number of touching spheres appears. 
Below the jamming threshold, the spheres do not touch. 
So one may conclude naively that there are no finite clusters below jamming. 
However, we may define appropriate non-trivial clusters in the following way. 
Consider a configuration of particles distributed on a one-dimensional system of length $L =Md_0$,  where $d_0$ is the diameter of the particles and $M$ for simplicity is an integer number. 
Divide then the length $L$ in $M$ segments of size $d_0$. 
Each particle will overlap with part of two adjacent segments, one overlap being smaller than the other. 
Then virtually shift the position of each particle to the centre of the segment where the overlap is larger. 
This procedure will lead to configurations of particles which map with one-dimensional percolation on a lattice. 
The clusters so defined will percolate at the jamming transition, with the same critical exponents of one-dimensional percolation on a lattice.

In conclusion, this example shows how to define clusters in the unjammed phase, in $1 d$. Extension in higher dimensions, as shown above, due to topological randomness, needs a more elaborate definition.

\section{Conclusions}
Our numerical investigation of the jamming transition in $d = 4$ and $d = 5$ spatial dimensions are consistent with previous results in $d=2$ and $d=3$. 
Taken together, all these results indicate that the approach to jamming can be described as random first-order percolation transition, with critical exponents $\beta =0$, $\gamma = 2$,
$\alpha = 0$ and finite size scaling exponent $\nu^* = 2/d$. for any $d \geq 2$. We have argued that the upper critical dimensionality for this mixed-order random percolation model is $d_u=2$ consistent with the common idea in the literature that $d_u=2$ is the upper critical dimension for jamming transition. Consequently the connectedness length exponent $\nu=1$ for any $d\geq 2$.

%The approach we have introduced to measure the characteristic cluster size, besides, has been inspired by that used to measure this length scale in glasses through the point-to-set method~\cite{biroli}.

In the numerical approach we have considered, the packing is prepared through the minimization of the energy of random arrangements of spheres, through the conjugate gradient method.
This protocol is one of the many different ones that could use to generate energy minima configurations.
Since different protocols lead to slightly different jamming volume fraction thresholds~\cite{pica_ciamarra_disordered_2010,Chaudhuri2010,pica_ciamarra_recent_2010}, they might in principle also affect the percolation scenario. \\
\\

{\bf Acknowledgements}
A.F. acknowledges financial support of the MIUR PRIN 2017WZFTZP ``Stochastic forecasting in complex systems''.
AP and MPC acknowledge support from the Singapore Ministry of Education through the Academic Research Fund (Tier 2) MOE2017-T2-1-066 (S) and from the National Research Foundation Singapore, and are grateful to the National Supercomputing Centre (NSCC) of Singapore for providing computational resources.

\end{document}